\documentclass{article}
\usepackage[utf8]{inputenc}
\usepackage{amsmath}
\usepackage{graphicx}
\usepackage{amssymb}
\usepackage{cite}
\usepackage{bm}
\usepackage{amsthm}
\usepackage{subcaption}
\usepackage{hyperref}
\usepackage{authblk}
\usepackage{caption}
\usepackage{subcaption}
\usepackage[toc,page]{appendix}
\usepackage[margin=1in]{geometry}

\usepackage[textwidth=1.7cm,textsize=small]{todonotes}

\title{Regularization of Single Field Inflation Models}

\author{Josh Hoffmann\thanks{Corresponding author: j.hoffmann1@lancaster.ac.uk} } 
\author{David Sloan}
\affil{Department of Physics, Lancaster University, Lancaster, United Kingdom LA1 4YB}

\date{}

\begin{document}

\maketitle

\begin{abstract}
    There are many single field inflationary models that are consistent with the recent Planck 2018 measurements of the spectral index $n_s$ and tensor-to-scalar ratio $r$. Despite good agreement with observational data some of these models suffer from having unregularized potentials which would produce a collapsing universe shortly after the end of inflation. In this paper we show that how one chooses to correct the behaviour potential towards the end of inflation can have a significant effect on the inflationary predictions of the model, specifically in the case of quartic hilltop and radiatively corrected Higgs inflation.
\end{abstract}

\section{Introduction}

Cosmological inflation represents one of the most important aspects of modern cosmology and is the focus of a great amount of theoretical and experimental physics \cite{Planck:2018nkj,Hinshaw2013NINEYEARWM, Lyth2007,PhysRevD.23.347, 1982PhLB..108..389L, Linde:2007fr}. Inflation is a period of exponential expansion of the early universe some time between $10^{-36}$ and $10^{-32}$ seconds after the big bang and is essential for understanding the structure of the CMB, a cornerstone of observational cosmology \cite{PhysRevD.23.347,1982PhLB..108..389L, Linde:2007fr, Vazquez:2018qdg}. The theory of cosmological inflation was originally posited as a means to explain several observations about our universe, namely the horizon, flatness and monopole problems \cite{Martin:2003bt}. However its true success is in providing a mechanism for structure formation in the early universe \cite{Martin2005, Martin:2007bw, PhysRevD.42.3413, Polarski:1995jg,Kiefer:1998qe,Kiefer:2008ku, Sudarsky:2009za, Martin:2012pea, Martin:2012ua, Peacock:1995qb, Linde:1982uu, PhysRevLett.49.1110, PhysRevD.28.679, liddle_lyth_2000}. Inflation allows for quantum mechanical density perturbations in the otherwise homogeneous matter content of the early universe to become amplified to large scales, eventually leading to the formation of structures we are more familiar with today such as stars, galaxies and the CMB. As such, it is important to provide an explanation as to how inflation occurs and to corroborate this with experimental observations.

It has been particularly fruitful to study the inflationary paradigm through the introduction of a homogeneous field $\phi(t)$ (typically a scalar) minimally coupled to gravity and governed by a potential $V(\phi)$ which essentially defines the model.
The Planck 2018 \cite{Planck:2018jri} survey has made it possible to precisely test a huge variety of inflationary models. Two important and model dependant cosmological observables are the scalar tensor ratio $r$ and spectral index $n_s$. These parameters and their compatibility with the parameter space determined by the Planck 2018 results are studied for many models in \cite{Martin:2013tda}.

The Planck 2018 survey estimates  $n_s = 0.9649 \pm 0.0042$ at $68\%$ CL \& $r\lesssim 0.056$ at $95\%$ CL. However a more recent BICEP/Keck array measurement \cite{BICEP:2021xfz} further constrained this to $r\lesssim 0.036$ at $95\%$ CL.

The spectral index $n_s$ and tensor fraction are directly related to the inflationary potential through the slow-roll parameters $\varepsilon(\phi)$ and $\eta(\phi)$ 

\begin{equation}
    n_s = 1 - 6\varepsilon + 2\eta
\end{equation}
\begin{equation}
    r = 16\varepsilon
\end{equation}

where

\begin{equation}
    \varepsilon(\phi) = \frac{1}{2}m_{pl}^2\left(\frac{V'(\phi)}{V(\phi)}\right)^2
\end{equation}
\begin{equation}
    \eta(\phi) = m_{pl}^2\frac{V''(\phi)}{V(\phi)}
\end{equation}
and thus allow us to directly test our inflationary models by specifying a form of the potential $V(\phi)$.

A number of the single scalar field models analysed in \cite{Martin:2013tda}, including some of those most favoured by the Planck survey, involve potentials which are unbounded from below after the end of inflation. Such models may be able to provide satisfactory inflationary dynamics, however if they are taken to be valid beyond inflation the potentials quickly become negative and due to their unboundedness, lead to universes that collapse on a time scale $\tau \sim H^{-1}$. 
Clearly this is not the cosmology we observe today, so such behaviour must be corrected by regularizing the potentials to introduce a local minima shortly after inflation ends. Naturally one may ask if modifying the behaviour after the end of inflation significantly changes the models predictions of inflationary observables such as the tensor fraction and spectral index, even 50-60 efolds before the end of inflation.

In this paper we explore the effect of generic correction terms on a collection of inflationary models. The correction terms are designed only to be the simplest possible options that prevent collapse of the universe after the end of inflation. Such terms which stabilise the inflaton VEV may, and do, appear in in more physically motivated in other physically motivated models.Although in this work we do not attempt to explain the physical origin of such terms. We show that such corrections will have an effect on the inflationary predictions of the model. Thus, regardless of how corrections of these forms may appear, they do indeed need to be accounted for before the inflationary predictions of the model can be trusted.

We show that the addition of correction terms designed to stabilise the inflaton VEV affect the reheating temperature $T_{re}$ predicted by the model as a function of the spectral index $n_s$ at a given number of efolds before the end of inflation $N_k$. Therefore by requiring that the predicted reheating temperature stays within the loose bounds set by the big bang nucleosynthesis scale and the energy scale of inflation, $0.01\text{GeV} \leq T_{re} \leq 10^{16}\text{GeV}$, the model under consideration may fall into a further restricted region of $r-n_s$ space that is acceptable under the Planck 2018 results. Thus we arrive at a key result of our analysis: In considering models which require regularization, it is insufficient to examine their predictions without taking such regularization into account. We have chosen to remain agnostic on the form that such regularization should take, motivating the correction terms solely by their role in ensuring that the potential remain positive with zero minimum. This allows us to demonstrate that generic regularizations should be taken into account regardless of their physical origins. In specific cases there are good physical reasons which inform the precise nature of the regularization, such as the contribution of higher loop corrections to the radiatively corrected Higgs inflation from the Jordan frame. However our goal is not simply to test specific models and regularizations but to make the broader point that since regularizations do make an impact on physical observables for such models, a wide range of potentials requiring regularization should only have their observational consequences examined with such regularizations in place.

In \cite{Hoffmann:2021vty} we addressed for the case of quartic hilltop inflation, a model well favoured by Planck, corrected by one particular modification scheme which involves squaring the potential. It is shown that modifying the quartic hilltop potential in this way does have a significant effect on the final $(r,n_s)$ parameter space that is consistent with the Planck 2018 results.
In this paper, we explore several different modification types for quartic hilltop, radiatively corrected Higgs inflation and Exponential SUSY inflation. These are all single field inflation models featured in \cite{Martin:2013tda} which all suffer from the same issue of unboundedness. We consider generic corrections to these models, motivated only by preventing collapse of the universe after the end of inflation, but we show that should such correction terms arise from physical motivations, they may have significant effects on the inflationary predictions of the models. Therefore one must carefully consider exactly how a candidate potential exits inflation. As we will show in this paper, it can have a significant effect on the inflationary predictions. Models that produce such a collapsing universe, must be corrected before calculations of $r$ and $n_s$ are compared to data. In all three models we explore the effects of simply squaring the potential to form a potential that is bounded below (for quartic hilltop this is investigated in detail in our previous paper \cite{Hoffmann:2021vty}). We then explore the simplest possible correction terms that may be used to extend the models, these are polynomial and inverse polynomial terms for QH and RCHI respectively.

The rest of this paper is structured as follows. In section 2 we will recap the quartic hilltop model (QH), a promising candidate potential. We summarise the current comparisons of the hilltop model predictions to the Planck 2018 survey. We also review recent analytical investigations of the model in \cite{DIMOPOULOS2020135688, Hoffmann:2021vty}. Polynomial corrections are then added to the hilltop model and their effects on inflationary observables are investigated. In section 3 we move on to the second model identified in Encyclopaedia Inflationaris \cite{Martin:2013tda}, radiatively corrected Higgs inflation (RCHI), which suffers from the same unboundedness issue as quartic hilltop. We explore a squared version of the potential as well as adding inverse polynomial correction terms to modify the behaviour of the potential towards the end of inflation, and their effects are investigated. It is certainly worth noting that the RCHI model derives from calculating quantum corrections to the Higgs inflation model in the Jordan frame and only then transforming to the Einstein frame potential. However there exists another method of consistently correcting the RCHI model, in which the quantum corrections are calculated in the Einstein frame \cite{Ferrara:2010in,Bezrukov:2010jz}, this will be discussed in more detail in section 3. In section 4 we discuss the final model, exponential SUSY inflation (ESI). ESI also suffers from the same unboundedness as QH and RCHI but is one of the most promising in terms of it's compatibility with Planck 2018. We find that using the approximation techniques in this paper, it is not possible to give the ESI potential a local mininma near where the field exits inflation at $\phi = 0$ by adding inverse polynomial terms. However, a squared version of this potential is investigated.

\section{Hilltop Models}
\subsection{Recap of Hilltop Inflation and Its Current State}

After the Planck 2018 survey, one of the most popular inflationary models that was consistent with the observational data was the Quartic Hilltop model. This model belongs to the wider class of hilltop models, those defined by a potential of the form 
\begin{equation}
\label{Hilltop}
    V(\phi) = V_0\left(1-\lambda\frac{\phi^n}{m_{pl}^n} + ...\right)
\end{equation}
The characteristic features of hilltop models make them very attractive for the slow-roll approximation. Namely that, in hilltop models contain a broad, flat plateau (or ``hilltop'') near $\phi = 0$. The two most popular cases are quadratic and quartic hilltop models, which have $n=2$ and $n=4$ respectively. Early analytical investigations of hilltop models \cite{LYTH19991} found that for $n>2$, provided the higher order terms are heavily suppressed during inflation, the spectral index  when the cosmological scale exits the horizon should be 
\begin{equation}
    n_s \simeq  1 - 2\left(\frac{n-1}{n-2}\right)\frac{1}{N}
\end{equation}
where $N$ is the remaining number of efolds of inflation after the scale $k_*$ exits the horizon. For the $n=4$  case of quartic hilltop inflation
\begin{equation}
\label{V_QH}
    V(\phi) = V_0\left[1-\lambda\left(\frac{\phi}{m_{pl}}\right)^4\right]
\end{equation}
the spectral index at $N = 50$ and $N = 60$ is $n_s = 0.94$ and $n_s = 0.95$ respectively. This is far too low considering the 2$\sigma$ Planck 2018 bound of $n_s \gtrsim 0.9607$

\begin{figure}[h]
    \centering
    \includegraphics[scale = 0.7]{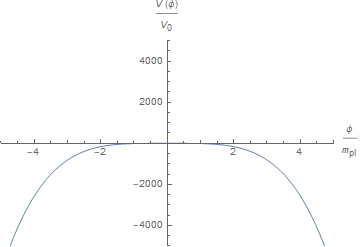}
    \caption{The quartic hilltop potential at $\lambda = 10$. This value if $\lambda$ is chosen only to clearly illustrate the shape of the potential, much smaller values of $\lambda$ are required to produce a spectral index and tensor fraction in agreement with the Planck 2018 measurements.}
    \label{V_QH_f}
\end{figure}

The $n=2$ case is simply that of Higgs inflation for $\phi \ll m_{pl}$ \cite{Rubio:2018ogq,Cheong:2021vdb} for which 
\begin{equation}
    n_s = 1-\frac{2}{N}, \quad r = \frac{8}{N}
\end{equation}
The spectral index at $N=50$ and $N = 60$ are within an acceptable range, however the tensor fraction 
is far larger than that allowed by the Planck 2018 bound $r \lesssim 0.1$. 

This initial analytical treatment of the hilltop potential was reliant on two key assumptions. Firstly that during inflation, the value of the scalar field is much smaller than it's vacuum expectation value $\phi_0 = m_{pl}\lambda^{-\frac{1}{4}}$. Secondly, in order to calculate the spectral index and tensor fraction as functions of $N$, one must first calculate 
\begin{equation}
\label{N_phi}
    N(\phi) = \frac{1}{m_{pl}}\int^{\phi}_{\phi_{\textrm{end}}} \frac{V(\phi)}{V'(\phi)} d\phi
\end{equation}
In the initial calculations in \cite{LYTH19991}, it's assumed that the contribution from the value of the inflaton field at the end of inflation is negligible. This is essentially a statement that the model can be assumed to exit inflation on the plateau where $V\simeq V_0$.

 The numerical analysis of these models carried out in \cite{Martin:2013tda} however revealed that such models can be compatible with observational data from the Planck 2013 survey if one considers the correct parameter range for $\lambda$. Namely $\lambda \ll 1$ such that the VEV is super Planckian $\langle \phi \rangle \gg m_{pl}$.
 In the Planck 2018 measurements of the spectral index and tensor fraction, the quartic hilltop (QH) model was featured as one of the more favourable models. \cite{Planck:2018jri}. The results from the extensive numerical investigations of the quartic hilltop model in \cite{Martin:2013tda} were still found to be compatible with the tighter constraints put on $n_s$ and $r$ from the 2018 measurements. However numerical simulations alone do not explain why the model behaves in this way, only the results that we should expect in such a parameter range. The behaviour of the QH model was made fully transparent by a recent analytical calculation of the spectral index and tensor fraction \cite{DIMOPOULOS2020135688}, in which the two previous assumptions are relaxed. When performing these calculations one finds specifically that the contribution of the inflation field at the end of inflation $\phi_{\text{end}}$ is
 \begin{equation}
     N_{\textrm{end}} = \frac{1}{4\sqrt{\lambda}}
 \end{equation}
 
 which is clearly large for sufficiently small $\lambda$. This analytical treatment of the quartic hilltop model made it possible to derive a closed form expression for the tensor fraction $r$ as a function of the spectral index $n_s$ and remaining number of efolds of inflation (after the scale exits the horizon) $N$.
 
 \begin{equation}
 \label{QH_r_ns}
     r(n_s,N) = \frac{8}{3}(1-n_s)\left[1-\frac{\sqrt{3[2(1-n_s)N-3]}}{(1-n_s)N}\right]
 \end{equation}
 Equation \ref{QH_r_ns} explains exactly the behaviour of the QH model in numerical simulations \cite{Martin:2013tda, CMB-S4:2016ple}. The relaxing of the assumptions of previous analytical investigations allows the value of the field during inflation to come closer to its VEV which is super-planckian in the small $\lambda$ regime, where the potential's plateau is broader and flatter allowing for more efolds of inflation to take place. We are also provided with an explanation for why the $\phi_{\text{end}}$ contribution cannot be disregarded in equation \ref{N_phi}, since this contribution is proportional to $\lambda^{-\frac{1}{2}}$. It's clear that in order for the QH model to be reconcile with observational data, we must take care in considering how the model exits inflation since $\phi(N)$ is allowed to come close to its VEV. Furthermore, it is well established that the potential \ref{V_QH} approximates a linear potential near the VEV, which is already known to be incompatible with the recent Planck data. Thus to accurately calculate the predictions of hilltop inflation, one has to account for the contribution of the vacuum stabilising terms in the potential which are suppressed during early inflation when $\phi(N)$ is small but become significant around the VEV.

 The importance of these terms is discussed extensively in \cite{Kallosh:2019jnl}. The authors consider general hilltop potentials of the form \ref{Hilltop} (reproduced here for brevity)
 \begin{equation}
     V(\phi) = V_0\left(1-\lambda\frac{\phi^n}{m_{pl}^n} + ...\right)
 \end{equation}
 
  stabilised by corrections, denoted ``...", which generically give the potential a local minima at $\phi_0 = \lambda^{-1/n}m_{pl}$. For such models, in the limit $\lambda \gg 1$, inflation ends before $\phi$ is near the VEV, and since $\phi \ll \phi_0$ for the duration of inflation, the ambiguity in the stabilising terms does not affect the predictions of $n_s$ and $r$, so it is suitable to disregard them in this limit. Of course, as we have outlined, the predictions of hilltop models in this limit are known to be poorly compatible with recent measurements, and the model is only a suitable fit in the $\lambda \ll 1$ regime where the behaviour of the potential around the VEV becomes significant. However the behaviour of the quartic hilltop model is vastly unphysical in this region. Since the QH potential is unbounded from below, after inflation ends, the potential soon becomes negative and the inflaton field has access to arbitrarily low energy states, and the universe would begin to collapse on a time scale $t\sim H^{-1}$. As the authors in \cite{Kallosh:2019jnl} point out, this is a generic result of potentials $V(\phi/\mu)$ which are unbounded from below, and such models should be excluded from the space of candidate inflationary potentials as they do not produce any sensible inflationary cosmology. Only those models with regularised potentials can be considered suitable candidates. Considering the QH model specifically, there are many ways of ensuring that the potential has a stable vacuum. The simplest options would just be to add terms which are higher order in $(\phi/m_{pl})^n$. One could also capture the effect of infinitely many such terms by simply taking the square of the QH potential as suggested in \cite{Kallosh:2019jnl}
\begin{equation}
    V(\phi) = V_0 \left[1-\lambda\left(\frac{\phi}{m_{pl}}\right)^4\right]^2
\end{equation}

\begin{figure}[h]
    \centering
    \includegraphics[scale = 0.7]{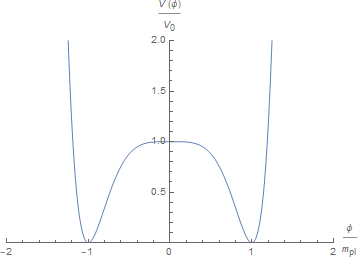}
    \caption{The quartic hilltop squared potential with $\lambda = 1$ for illustrative purposes.}
    \label{V_QHS}
\end{figure}

For small $\phi$ this potential is $V(\phi) \simeq V_0(1-2\lambda\phi^4/m_{pl}^4)$, thus the behaviour is only modified near the VEV. This ``Quartic Hilltop-Squared" (QHS) potential has been investigated numerically in \cite{Kallosh:2019jnl}, and like it's quartic hilltop cousin, was found to be in good agreement with the Planck 2018 constraints on $r$ and $n_s$, but with the added benefit of producing sensible, and physically viable inflationary cosmology due to the vacuum stabilising terms contained in it's series expansion.  
 
Accounting for these terms also makes the potential suitable for analysing reheating after inflation. Which is ultimately necessary as the inflationary predictions of the model must be compatible with it's predictions of the temperature at the end of reheating $T_{re}$ and the number of efolds of reheating $N_{re}$ which constrain $N$.

\subsection{Corrected Hilltop Models}
While the predictions of the tensor fraction and spectral index for the QHS model are consistent with the Planck 2018 bounds, the numerical analysis that these results are derived from still acts in an opaque manner. It tells us only the results that we should expect for $r(n_s)$ and not precisely why the model behaves in a certain way. Further more, this analysis is incomplete with regards to reheating after inflation. Since the QHS potential is UV regulated, there exists a local minima of the potential just after inflation ends around which the inflaton oscillate. The damped oscillations cause the inflaton field to dump it's energy into the thermal bath of the universe, leading eventually the production of ordinary matter. The temperature of the universe at the end of reheating $T_{re}$ and the number of reheating efolds $N_{re}$ depend explicitely on the number of efolds of inflation after the scale exits $N_k$.
\begin{equation}
    \label{T_re}
        T_{re} = V_0\left[\left(\frac{43}{g_{re}}\right)^{\frac{1}{3}}\left(\frac{a_0T_0}{k}\right)H_k e^{-N_k}\left(\frac{45}{\pi^2g_{re}}\right)^{-\frac{1}{3(1+w_{re})}}\right]^{\frac{3(1+w_{re})}{3w_{re}-1}} 
\end{equation}
\begin{equation}
    \label{N_re}
    N_{re} = \frac{4}{1-3w_{re}}\left(\frac{1}{4}\ln\frac{\pi^2g_{re}}{45}+\frac{1}{3}\ln\frac{11}{43g_{re}}+\ln\frac{a_0T_0}{k}-\ln\frac{V_{end}^{\frac{1}{4}}}{H_k}-N_k\right)
\end{equation}
where $g_{re}$ is the effective number of relativistic species at the end of reheating, $T_0$ is the present day CMB temperature, $V_{end}$ is the value of the inflaton potential at the end of inflation, $H_k$ is the value of the Hubble parameter when the scale exits and $w_{re}$ is the effective equation of state parameter of the inflaton during reheating. Throughout this paper we will assume a mean reheating parameter of $w_{re} = 1$ as this providers the most conservative estimates on the predictions of reheating \cite{Munoz:2014eqa}.

For a fixed value of $N_k$, we see that the $(r,n_s)$ predictions of the QH model differ quite drastically from the QHS model, this behaviour can be better understood by also preforming an analytical investigation of the QHS model. The analytical methods developed in \cite{DIMOPOULOS2020135688} to investigate the quartic hilltop model have also been used in \cite{Hoffmann:2021vty} analyse the inflationary predictions of QHS. In that analysis we derive a closed form expression for $r(n_s)$
\begin{equation}
\label{r}
    r = \begin{cases}
    \frac{1}{4N}\left(\frac{12N\tilde{n}_s-2N^2\tilde{n}_s^2-15 + \sqrt{8N\tilde{n}_s-15}}{4N\tilde{n}_s-N^2\tilde{n}_s^2+\sqrt{8N\tilde{n}_s-15}}\right)\left[1+g_+(\tilde{n}_s, N)\right]\left[g_+(\tilde{n}_s, N)-7\right], \quad \lambda \geq \lambda_c \\
    \\
    
    \frac{1}{4N}\left(\frac{12N\tilde{n}_s-2N^2\tilde{n}_s^2-15 - \sqrt{8N\tilde{n}_s-15}}{4N\tilde{n}_s-N^2\tilde{n}_s^2 - \sqrt{8N\tilde{n}_s-15}}\right)\left[1+g_-(\tilde{n}_s, N)\right]\left[g_-(\tilde{n}_s, N)-7\right], \quad \lambda < \lambda_c
    \end{cases}
\end{equation}

where 
\begin{equation}
    g_\pm(\tilde n_s, N) = \sqrt{\frac{12N\tilde n_s+62N^2\tilde n_s^2-16N^3\tilde n_s^3-15\pm(1+16N\tilde n_s)\sqrt{8N\tilde n_s-15}}{12N\tilde n_s-2N^2\tilde n_s^2-15 \pm \sqrt{8N\tilde n_s-15}}}
\end{equation}
and the critical parameter value is $\lambda_c = (60N)^{-\frac{1}{2}}$. As noted previously, in the QH model, there is a contribution $N_{end}^{QH} = 1/4\sqrt{\lambda}$ to the total efolds of inflation coming from the value of the field at the end of inflation $\phi_{end}$. In the QHS model this contribution is found to be $N_{end}^{QHS} = 1/8\sqrt{\lambda}$ (when $\lambda\ll1$) the contribution is half as small as in the QHS model, for small $\lambda$ the higher order terms begin to contribute significantly, and push the field out of slow roll inflation sooner than in QH. For all of the corrected quartic hilltop models in this paper, we work in the small $\lambda$ regime, where the symmetry breaking scale $\mu = m_{pl}(4\lambda)^{-1/4}$ is large \cite{Kinney:1995cc}, thus avoiding the region of parameter space that is not compatible with the Planck results.

Furthermore, the authors complete the analysis of QHS model by computing it's reheating temperature and efolds. Computing the reheating parameters allows us to further constrain the parameter space for QHS by demanding that the reheating temperature be bounded below by the Big Bang Nucleosynthesis scale $T_{BBN} \lesssim 0.01\text{GeV} \lesssim T_{re}$ and bounded above by the energy scale of inflation $T_{re}\lesssim 10^{16}\text{GeV}$. In fact, the size of the acceptable region in $(r,n_s)$ space of the QHS model reduces significantly after reheating consistency is taken into account. Before reheating, whilst only bounded by the Planck 2018 data, the QHS model was constrained to the region defined by 
\begin{equation}
    \begin{aligned}
    0.9607 \lesssim &n_s \lesssim 0.9691 \\
    55 \lesssim &N_k
    \end{aligned}
\end{equation}
whereas after taking into account reheating consistency, this is reduced to 
\begin{equation}
    \begin{aligned}
        0.9607 \lesssim &n_s \lesssim 0.9691 \\
        63 \lesssim &N_k \lesssim 68
    \end{aligned}
\end{equation}
The fact that the inflationary potential must be UV regularised and that the models predictions of the tensor fraction and spectral index must be consistent with reheating bounds significantly constrains the models parameter space which is in agreement with the Planck 2018 bounds.

Choosing to regularise the potential by squaring it, is simply one possibility out of a great many. In this work, we show that how one chooses the regularize the shape of the potential near the end of inflation can have an important and quantifiable effect on the models predictions of $n_s$ and $r$. The QHS potential essentially amounts to adding a $(\phi/m_{pl})^8$ correction term to the original $QH$ potential, which is seen simply by expanding out the potential and rescaling the coupling parameter $\lambda$.
\begin{equation}
        V_\textrm{QH}(\phi) = V_0\left[1-\lambda\left(\frac{\phi}{m_{pl}}\right)^4\right]^2 
\end{equation}
\begin{equation}
\label{QHS_8}
    \begin{aligned}
        V_\textrm{QHS}(\phi) &= V_0\left[1-2\lambda\left(\frac{\phi}{m_{pl}}\right)^4 + \lambda^2\left(\frac{\phi}{m_{pl}}\right)^8 \right] \\
        &= V_0\left[1-\tilde{\lambda}\left(\frac{\phi}{m_{pl}}\right)^4+\frac{1}{4}\tilde{\lambda}^2\left(\frac{\phi}{m_{pl}}\right)^8\right]
    \end{aligned}
\end{equation}

As well as squaring the potential, one may also consider what the effect of adding single polynomial terms $\phi^p$ to the potential are. We only wish to correct the behaviour of the potential for large $\phi$, so we look at positive values of $p$. Furthermore, the coefficient of the $\phi^p$ term must be fine-tuned to ensure that the VEV remains at $V(\phi_0) = 0$ so that inflation ends in a finite time. Therefore consider potentials of the form

\begin{equation}
    V(\phi) = V_0\left[1-\lambda\left(\frac{\phi}{m_{pl}}\right)^4 + \alpha_p\left(\frac{\phi}{m_{pl}}\right)^p\right], \quad p > 4
\end{equation}

The potential has stationary points at 
\begin{equation}
    \phi^3\left(p\alpha_p\phi^{p-4}-4\lambda\right) = 0
\end{equation}

and thus the VEV is located at 
\begin{equation}
    \phi_0 = \left(\frac{4\lambda}{p\alpha_p}\right)^{\frac{1}{p-4}}
\end{equation}
Requiring that $V(\phi_0) = 0$ allows us to solve for the fine-tuning of the coefficient $\alpha_p$
\begin{equation}
    \alpha_p = 4\lambda^{\frac{p}{4}}p^{-\frac{p}{4}}\left(p-4\right)^{\frac{p-4}{4}}
\end{equation}
So to add polynomial corrections to the Quartic Hilltop model, we work with potentials of the form
\begin{equation}
    V(\phi) = V_0\left[1-\lambda\left(\frac{\phi}{m_{pl}}\right)^4 + 4\lambda^{\frac{p}{4}}p^{-\frac{p}{4}}\left(p-4\right)^{\frac{p-4}{4}}\left(\frac{\phi}{m_{pl}}\right)^p\right]
\end{equation}
Henceforth we will refer to such models as $\text{QH}_p$, with the quartic hilltop squared model being equivalent to $\text{QH}_8$ as per equation \ref{QHS_8}.

The lowest order term that we can add to modify the small-field behaviour of the potential but retain its plateau and large-field shape is $p=5$ since the QH model already contains a $\phi^4$ term. As we shall see from the results later in this section, there is no need to investigate larger than $p=10$, so we consider only terms in this range. Starting from $p=5$ through to $p = 10$  we have calculated the tensor ratio $r$ as a function of the spectral index $n_s$ and the temperature at the end of reheating $T_{re}$ as a function of the spectral index between $50\leq N_k \leq N_k = 70$. We then determine which of these curves are reheating-consistent by demanding that they lie within the rectangle on the $T_{re}-n_s$ plots. The horizontal bounds on the reheating temperature plots are $0.01 \text{GeV} \lesssim T_{re} \lesssim 10^{16} \text{GeV}$. The reheating temperature is bounded below by the energy scale of Big Bang Nucleosynthesis (BBN) \cite{Cook:2015vqa,Lozanov:2019jxc} $T_{BBN} \sim 10\text{MeV}$. Measurements of the CMB anisotropies constraining the tensor fraction $r$ are equivalent to upper bounds on the energy scale of inflation \cite{Planck:2018jri, Lozanov:2019jxc,challinor_2012} since 
\begin{equation}
    V_*^{\frac{1}{4}} = \left(\frac{3\pi^2A_s}{2}rm_{pl}\right)^{\frac{1}{4}} \lesssim 10^{16}\text{GeV}
\end{equation}
where $V_*$ is the energy scale of inflation and $A_s$ is the amplitude of scalar perturbations. This bounds $T_{re}$ from above. 

In all of the following $r-n_s$ plots, we produce results from numerical solutions of the tensor fraction and spectral index and the corresponding temperature at the end of reheating for $\text{QH}_p$ models with $p = 5,10$. Figures for $p=6,7,9$ are contained in Appendix B (omitting $\text{QH}_8 \equiv \text{QHS}$). The solid and dashed black curves are the Planck 2018 $1\sigma$ and $2\sigma$ bounds respectively. The horizontal black line is the BICEP/Keck Array bound on the tensor fraction $r< 0.036$ \cite{challinor_2012}.
Similarly, in all following $T_{re}-n_s$ plots, the upper horizontal line is the energy scale of inflation bound $T_{re} \lesssim 10^{16}\text{GeV}$ and the lower horizontal line is the BBN bound $0.01\text{GeV} \lesssim T_{re}$. The vertical solid lines represent the Planck 2018 bounds $0.9607\lesssim n_s \lesssim 0.9691$.

\begin{figure}[h]
  \centering
  \begin{subfigure}{.45\linewidth}
    \centering
    \includegraphics[width = \linewidth]{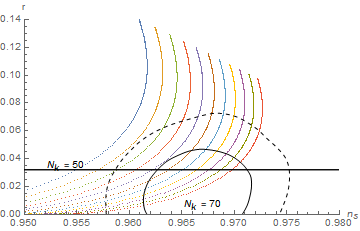}
    \caption{Numerical solutions of the tensor-scalar ratio $r$ and spectral index $n_s$ for the $\text{QH}_5$ model over the $50 \leq N_k \leq 70$.}
    \label{QH_5_rns}
  \end{subfigure}%
  \hspace{3em}%
  \begin{subfigure}{.45\linewidth}
    
    \centering
    \includegraphics[width = \linewidth]{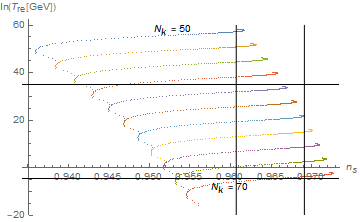}
    \caption{The (log) reheating temperature of the $\text{QH}_5$ model against the spectral index $n_s$ over $50\leq N_k \leq 70$.}
    \label{QH_5_Tns}
  \end{subfigure}%
  \caption{}
\end{figure}

Starting with the $\text{QH}_5$ model, only those curves with $58\lesssim N_k \lesssim 68$ are within the acceptable region of reheating temperate and spectral index in figure \ref{QH_5_Tns}. However when one also takes into account the $1\sigma$ region in figure \ref{QH_5_rns} this is further reduced to $60\lesssim N_k \lesssim 68$. Requiring that we consider both the Planck-consistent and reheating-consistent curves thus drastically reduces the acceptable region of $r-n_s$ parameter space for the corrected quartic hilltop model. This is significant as the corrections themselves are required for such models to even be taken seriously as candidates for single-field slow-roll inflation as without regularizing the potentials they produce a cosmology entirely incompatible with a universe that does not immediately collapse after inflation ends.

For $\text{QH}_5$, the resulting acceptable region on the $r-n_s$ plot lies in the mid to upper-left corner of the $1\sigma$ bound. If further, more precise measurements were to constrain the tensor fraction and spectral index away from this region, $\text{QH}_5$ would quickly become untenable.

\begin{figure}[h]
  \centering
  \begin{subfigure}{.45\linewidth}
    \centering
    \includegraphics[width = \linewidth]{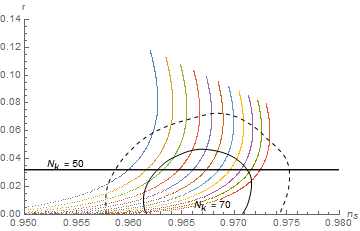}
    \caption{Numerical solutions of the tensor-scalar ratio $r$ and spectral index $n_s$ for the $\text{QH}_{10}$ model over the $50 \leq N_k \leq 70$.}
    \label{QH_10_rns}
  \end{subfigure}%
  \hspace{3em}%
  \begin{subfigure}{.45\linewidth}
    \centering
    \includegraphics[width = \linewidth]{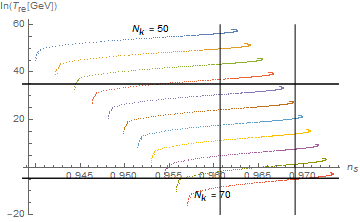}
    \caption{The (log) reheating temperature of the $\text{QH}_{10}$ model against the spectral index $n_s$ over $50\leq N_k \leq 70$.}
    \label{QH_10_Tns}
  \end{subfigure}%
  \caption{}
\end{figure}

Considering the $\text{QH}_{10}$ in figure \ref{QH_10_rns}, we see that increasing $p$ causes more curves to enter the $1\sigma$ region of the $r-n_s$ plot. If one were to not consider reheating consistency this would place a lower bound of $56 \lesssim N_k$ on the number of efolds of inflation after the scale exits. However as we increase the power of the correction term, the curves in the (log) reheating temperature plot move very slowly and by $p = 10$ no new curves have entered or left the acceptable region in figure \ref{QH_10_Tns}. Reheating consistency of $\text{QH}_{10}$ thus still requires $60 \lesssim N_k \lesssim 68$ just as with the lowest order correction $\text{QH}_5$. In contrast to $\text{QH}_5$ we see that now the curves for which $54 \lesssim N_k \lesssim 58$ have now entered the $1\sigma$ region in figure \ref{QH_10_rns} however must still reject these as they remain outside the acceptable region of the reheating plot.

\begin{figure}[h]
  \centering
  \begin{subfigure}{.45\linewidth}
    \centering
    \includegraphics[width = \linewidth]{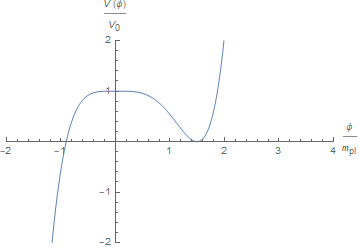}
    \caption{The $\text{QH}_5$ potential at $\lambda = 1$}
    \label{V_QH5}
  \end{subfigure}%
  \hspace{3em}%
  \begin{subfigure}{.45\linewidth}
    \centering
    \includegraphics[width = \linewidth]{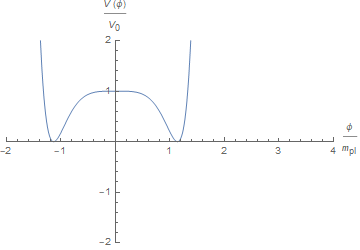}
    \caption{The $\text{QH}_{10}$ potential at $\lambda = 1$}
    \label{V_QH10}
  \end{subfigure}%
  \caption{}
\end{figure}

In figures \ref{V_QH5} and \ref{V_QH10} we plot the potentials of two of the corrected hilltop models $\text{QH}_5$ and $\text{QH}_{10}$ for $\lambda = 1$. The potentials now have a stable vacuum about which the field can oscillate during the reheating phase.

Overall, the region of reheating-consistent parameter space in the $r-n_s$ plot actually increases with increasing $p$, as the distance between the curves remains relatively constant and they are only shifted along laterally. Initially it may then seem that one could simply add arbitrarily high powers to the quartic hillop model in order to increase the region of validity.

\section{Radiatively Corrected Higgs Inflation}
The case of the quartic hilltop model demonstrates clearly that we must only consider those potentials that are regularised and will not collapse the universe immediately after exiting inflation. How one regularises the potential is important if the model is to be compared to measurements. A huge number of candidate single field inflation models are analysed in \cite{Martin:2013tda}. Out of these models there are two which suffer from the same vacuum stabilisation as quartic hilltop. The first of these is the Radiatively Corrected Higgs Inflation (RCHI) model. This model derives from taking into account 1-loop corrections to the Higgs inflation model, in which the inflaton is a Higgs particle \cite{Rubio:2018ogq, PhysRevD.52.4295, KAISER199423,vanderBij:1994bv,vanderBij:1993hx}. If one considers the standard model with the Higgs non-minimally coupled to classical gravity, the simplest such action in the Jordan frame is
\begin{equation}
    S_T = \int d^4x\sqrt{-g}\left[\frac{1}{2}m_{pl}^2R + \xi H^\dagger H R + \mathcal{L}_{SM}\right]
\end{equation}
where $H^\dagger$ is the Higgs doublet and $\xi > 0$ is a dimensionless coupling parameter.

If we consider only the gravi-scalar sector of the theory and make use of the unitary gauge for which $H = (0,h/\sqrt{2})$ where $h$ is a real scalar field, then the action of interest is 
\begin{equation} 
\label{S_Jordan}
    S = \int d^4x\sqrt{-g}\left[\frac{1}{2}F(h)R - \frac{1}{2}\partial_\mu h \partial^\mu h - W(h)\right]
\end{equation}
the functions $F(h)$ and $W(h)$ which we define now for later convenience are
\begin{equation}
    F_1(h) = (m_{pl}^2 + \xi h^2), \quad W(h) = \frac{\lambda}{4}(h^2-\nu^2)
\end{equation}

where $\nu$ is the elctroweak symmetry breaking scale. After a conformal transformation of the metric 

\begin{equation}
    g_{\mu\nu} \rightarrow \theta(x) g_{\mu\nu}, \quad \theta(x) = \frac{m_{pl}^2}{m_{pl}^2 + \xi h^2}
\end{equation}

we obtain the Einstein frame action 

\begin{equation}
    S = \int d^4x \sqrt{-g}\left[\frac{1}{2}m_{pl}^2R - \frac{1}{2}m_{pl}^2 K(\theta)\partial_\mu \theta \partial^\mu \theta - V(\theta)\right]
\end{equation}
 in which the scalar field $\theta$ has a non-trivial kinetic term 
 \begin{equation}
     K(\theta) = \frac{1}{4|a|\theta^2}\left(\frac{1-6|a|\theta}{1-\theta}\right), \quad a = -\frac{\xi}{1+6\xi}
 \end{equation}
 After rescaling to a canonically normalised field the action becomes 
 \begin{equation}
     S = \int d^4x \sqrt{-g} \left[\frac{1}{2}m_{pl}^2R - \frac{1}{2}\partial_\mu\phi\partial^\mu\phi - V(\phi)\right]
 \end{equation}
where the field $\phi$ satisfied the differential equation

\begin{equation}
    \frac{1}{M_{pl}^2}\left(\frac{d\phi}{d\theta}\right)^2 = K(\theta)
\end{equation}
The exact solution, which we omit here for brevity, is given in \cite{Rubio:2018ogq}.
At tree level, where the coupling in the Jordan frame $\xi$ is large and $a\simeq 1/6$ the Einstein frame potential is approximately

\begin{equation}
\label{V_Higgs}
    V(\phi) = \frac{M_{pl}^4\lambda}{4\xi^2} \left(1-e^{-\sqrt{\frac{2}{3}}\frac{\phi}{m_{pl}}}\right)^2
\end{equation}

We see from equation \ref{V_Higgs} that the Higgs self coupling $\lambda$ and coupling to gravity $\xi$ only enter the potential through it's overall normalisation so the ratio $\sqrt{\lambda}/\xi$ is completely determined from the CMB normalisation. However, the 1-loop radiative corrections to the effective action \ref{S_Jordan} in the Jordan frame contribute more significantly at large values of $\xi$, so it is not sufficient to consider only the tree level approximation \cite{Barvinsky:2008ia}. The radiative corrections modify the functions $F(h)$ and $W(h)$. To first order these corrections are 
\begin{equation}
\label{F}
    F(h) = m_{pl}^2 + \xi h^2 + \frac{C}{16\pi^2}h^2\ln\left(\frac{m_{pl}^2h^2}{\mu^2}\right)
\end{equation}

\begin{equation}
\label{W}
    W(h) = \frac{1}{4}\lambda\left(h^2-\nu^2\right)^2 + \frac{\lambda A}{128\pi^2}h^4\ln(\frac{h^2}{\mu^2})
\end{equation}

The modifications to these functions change the resulting Einstein frame potential from which we compute the inflationary observable $n_s$ and $r$. The Radiatively Corrected Higgs Inflation (RCHI) potential is thus 
\begin{equation}
\label{RCHI_V}
    V(\phi) \simeq V_0 \left(1-2e^{-\frac{2\phi}{\sqrt{6}m_{pl}}} + \frac{A_I}{16\pi^2}\frac{\phi}{\sqrt{6}m_{pl}}\right)
\end{equation}
where $A_I$ is a free parameter as given in eq 4.12 of \cite{Martin:2013tda}.

Before moving on to calculating corrections terms to the RCHI model we take note that there exists in the literature a modification to the Higgs inflation model that produces a regularized potential (as opposed to equation \ref{RCHI_V}) which is unbounded from below. Firstly, in the literature there has been debate over the validity of the Higgs inflation model due to it's UV cutoff being very close to the energy scale of inflation $H \sim \sqrt{\lambda} m_{pl}/\xi$ \cite{Ferrara:2010in,Bezrukov:2010jz,Barbon:2009ya, Park:2008hz, Germani:2010gm } The cutoff scale, at which unitarity is broken for tree-level amplitudes, is calculated in \cite{Bezrukov:2010jz} as $\Lambda \sim m_{pl}/\xi$ in the Einstein frame. For non-minimal coupling $\xi$ and $\lambda \sim O(1)$ the effective field theory may not be a valid description of inflationary dynamics. In \cite{Ferrara:2010in} the authors point out that these calculations of the cutoff scale are performed in the small field approximation $\phi \approx h$. However the authors note that the inflationary regime has $\phi \gg 1$, in which the Einstein frame potential is

\begin{equation}
\label{RCHI_Reg}
    \begin{aligned}
        V(\phi) & \approx \frac{\lambda}{4\xi^2}\left(1+e^{-\frac{2\phi}{\sqrt{6}}}\right)^2 \\
        &\approx \frac{\lambda}{4\xi^2} \left(1+2e^{-\frac{2\phi}{\sqrt{6}}}\right)
    \end{aligned}
\end{equation}

If one considers the series expansion of the potential in powers of $\phi$, the result is quite different to what would have been obtained in the small field approximation

\begin{equation}
    V(\phi) \approx \frac{3\lambda}{4\xi^2} - \frac{\lambda}{\xi^2\sqrt{6}}\phi + \frac{\lambda}{6\xi^2}\phi^2 - \frac{\lambda}{9\sqrt{6}\xi^2}\phi^3 +...
\end{equation}
thus the one-loop quantum corrections will contribute a cutoff 
\begin{equation}
    \Lambda \sim \frac{\xi^2}{\lambda}m_{pl}
\end{equation}
This unitarity bound is well above the energy scale of inflation and so the effective field theory remains a safe description of inflationary physics with a regularized potential given by equation \ref{RCHI_Reg}. In this paper however we focus only on the RCHI model featured in \cite{Martin:2013tda} as a means of demonstrating the effect of correction terms required by such unregularized potentials.

The Radiatively Corrected Higgs Inflation model shares similarities to Quartic Hilltop in that it also fits within the $1\sigma$ region of the Planck 2018 data over a given range of it's free parameter $A_I$, but does not produce a cosmology consistent with a universe that does not collapse

\begin{figure}[h]
  \centering
  \begin{subfigure}{.45\linewidth}
    \centering
    \includegraphics[width = \linewidth]{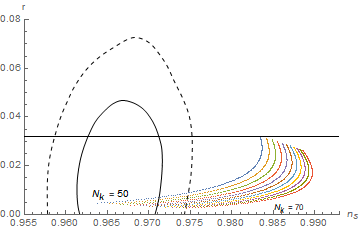}
    \caption{Numerical solutions for tensor fraction $r$ and spectral index $n_s$ of the RCHI model over the range $1\leq A_I \leq 40$ for $50\leq N_k \leq 70$. The solid and dashed curves represent the Planck 2018 $1\sigma$ and $2\sigma$ bounds respectively. The horizontal black line is the BICEP tensor fraction bound $r\lesssim 0.032$.}
    \label{RCHI_rns}
  \end{subfigure}%
  \hspace{3em}%
  \begin{subfigure}{.45\linewidth}
    \centering
    \includegraphics[width = \linewidth]{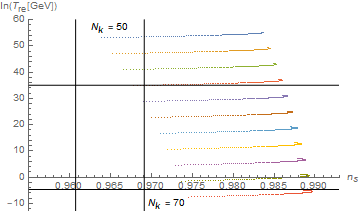}
    \caption{Numerical solutions for reheating temperature $T_{re}$ and spectral index $n_s$ of the RCHI model over the range $1\leq A_I \leq 40$ for $50\leq N_k \leq 70$. The vertical solid lines represent the Planck 2018 spectral index bounds $0.9607\lesssim n_s \lesssim 0.9691$ and the horizontal solid lines represent the reheating temperature bounds $0.01\text{GeV} \lesssim T_{re} \lesssim 10^{16}\text{GeV}$.}
    \label{RCHI_Tns}
  \end{subfigure}%
  \caption{}
\end{figure}

In figure \ref{RCHI_rns} that there are indeed a few curves that fit well within the $1\sigma$ region provided by the Planck 2018 data, specifically those with $50 \lesssim N_k \lesssim 60$. When we take into account reheating consistency over the same parameter range in figure \ref{RCHI_Tns} we see that lie within the acceptable region formed by the rectangular bounds, essentially ruling out this model as a possibility. Like the quartic hilltop model, this potential also fails to produce a cosmology compatible with what we observe today. In the RCHI model, we exit inflation at small $\phi$ and shortly after the potential becomes negative. This behaviour of the potential needs to be corrected. There are a variety of ways one may correct the behaviour of this potential, as we did to form the QHS model. As such we could first consider simply squaring the potential to produce one which is regularised. However for RCHI, the potential only needs to be corrected at small field values as we would like to retain the flatness of the potential during slow-roll. Squaring the potential does not achieve this very well as it makes the potential very steep at large $\phi$.

\begin{figure}[h]
  \centering
  \begin{subfigure}{.45\linewidth}
    \centering
    \includegraphics[width = \linewidth]{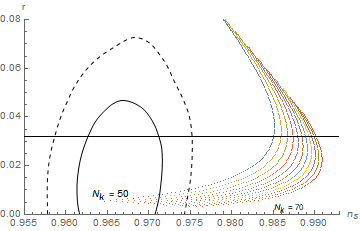}
    \caption{Numerical solutions for tensor fraction $r$ and spectral index $n_s$ of the RCHI-Squared model over the range $1\leq A_I \leq 40$ for $50\leq N_k \leq 70$. The solid and dashed curves represent the Planck 2018 $1\sigma$ and $2\sigma$ bounds respectively. The horizontal black line is the BICEP tensor fraction bound $r\lesssim 0.032$}
    \label{RCHI_Squared_rns}
  \end{subfigure}%
  \hspace{3em}%
  \begin{subfigure}{.45\linewidth}
    \centering
    \includegraphics[width = \linewidth]{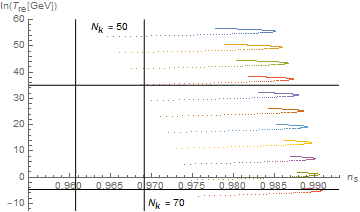}
    \caption{Numerical solutions for reheating temperature $T_{re}$ and spectral index $n_s$ of the RCHI model over the range $1\leq A_I \leq 40$ for $50\leq N_k \leq 70$. The vertical solid lines represent the Planck 2018 spectral index bounds $0.9607\lesssim n_s \lesssim 0.9691$ and the horizontal solid lines represent the reheating temperature bounds $0.01\text{GeV} \lesssim T_{re} \lesssim 10^{16}\text{GeV}$.}
    \label{RCHI_Squared_Tns}
  \end{subfigure}%
  \caption{}
\end{figure}

Compared to the standard RCHI model, the $r-n_s$ curves in figure \ref{RCHI_Squared_rns} are elongated and far more dramatically curved outside of the $2\sigma$ region. Whilst the curves with $50 \lesssim N_k \lesssim 58$ lie within the $1\sigma$ region, it still remains the case that none of the curves are able to be made reheating-consistent. Furthermore, it is also worth noting that one needs at least 60 efolds of inflation to solve the horizon problem, making a squared version of the RCHI potential even less favourable.

\begin{figure}[h]
  \centering
  \begin{subfigure}{.45\linewidth}
    \centering
    \includegraphics[width = \linewidth]{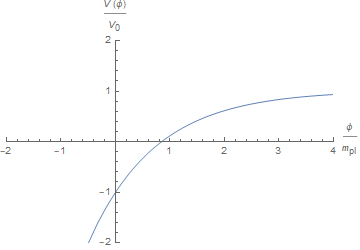}
    \caption{The RCHI potential at $A_I = 1$}
    \label{V_RCHI}
  \end{subfigure}%
  \hspace{3em}%
  \begin{subfigure}{.45\linewidth}
    \centering
    \includegraphics[width = \linewidth]{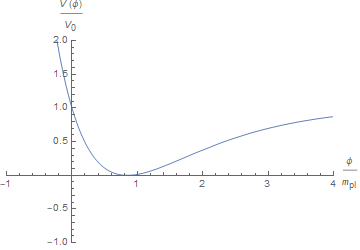}
    \caption{The RCHI-Squared potential at $A_I = 1$.}
    \label{V_RCHIS}
  \end{subfigure}%
  \caption{}
\end{figure}

The simplest such terms that one could add the only modify the potential at small $\phi$ are inverse powers of the form $\phi^{-p}$ where $p>0$. Henceforth we refer to such models as $\text{RCHI}_p$. Just as we did when adding polynomial correction to the quartic hilltop model, we must ensure that the coefficient of the correction term keeps the scalar fields VEV at zero in order for inflation to end in a finite time. That is, consider a potential of the form 
\begin{equation}
    V(\phi) = 1 - 2e^{-\frac{2\phi}{\sqrt{6}m_{pl}}} + \frac{A_I}{16\pi^2}\frac{\phi}{\sqrt{6}m_{pl}} + \alpha_p6^{\frac{p}{2}}\phi^{-p}
\end{equation}

The coefficient $\alpha_p$ must be such that $V(\phi_0) = 0$, where $\phi_0$ is the field value at the minimum of the potential and therefore satisfies $V'(\phi_0) = 0$

\begin{equation}
\label{p_0_exact}
    \frac{4\phi_0}{\sqrt{6}m_{pl}}e^{-\frac{2\phi_0}{\sqrt{6}m_{pl}}} + \frac{A_I}{16\pi^2\sqrt{6}m_{pl}} -p\alpha_p 6^{\frac{p}{2}}\phi_0^{-(p+1)} = 0
\end{equation}

Equation \ref{p_0_exact} cannot be solved analytically for $\phi_0$ but we may make use of an approximation. The minima of the potential is at small field values $\phi_0/m_{pl} \lesssim  1$ and thus 
\begin{equation}
    e^{-\frac{2\phi_0}{\sqrt{6}m_{pl}}} \approx 1 - \frac{2\phi_0}{\sqrt{6}m_{pl}}
\end{equation}
The $\phi_0^{-(p+1)}$ term will dominate the expression and thus we may solve for $\phi_0(\alpha_p)$
\begin{equation}
    \label{p_0_app}
    \phi_0 \simeq \sqrt{6}\left(\frac{16\pi^2p\alpha_p m_{pl}}{A_I + 64\pi^2}\right)^{\frac{1}{p+1}}
\end{equation}

Now substituting approximate solution \ref{p_0_app} into $V(\phi_0) = 0$ allows us to solve for the coefficient 
\begin{equation}
\label{a_RCHI}
    \alpha_p = \left(\frac{16\pi^2m_{pl}}{A_I + 64\pi^2}\right)^p p^p (1+p)^{-(p+1)}
\end{equation}
and so we work with an approximate potential 
\begin{equation}
    V(\phi) = 1 - 2e^{-\frac{2\phi}{\sqrt{6}m_{pl}}} + \frac{A_I}{16\pi^2}\frac{\phi}{\sqrt{6}m_{pl}} + \left(\frac{16\pi^2m_{pl}}{A_I + 64\pi^2}\right)^p p^p (1+p)^{-(p+1)} 6^{\frac{p}{2}}\phi^{-p}
\end{equation}

In the following figures we display the $r-n_s$ and $T_{re}-n_s$ plots for $\text{RCHI}_p$ with $1\leq p \leq 10$. The bounds on the parameter spaces are the same as in all previous figures.

\begin{figure}[h]
  \centering
  \begin{subfigure}{.45\linewidth}
    \centering
    \includegraphics[width = \linewidth]{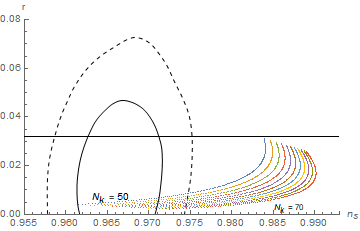}
    \caption{Numerical solutions of the tensor-scalar ratio $r$ and spectral index $n_s$ for the $\text{RCHI}_1$ model over the $50 \leq N_k \leq 70$.}
    \label{RCHI_1_rns}
  \end{subfigure}%
  \hspace{3em}%
  \begin{subfigure}{.45\linewidth}
    \centering
    \includegraphics[width = \linewidth]{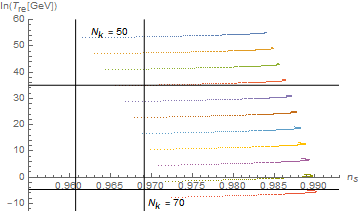}
    \caption{The (log) reheating temperature of the $\text{RCHI}_1$ model against the spectral index $n_s$ over $50\leq N_k \leq 70$.}
    \label{RCHI_1_Tns}
  \end{subfigure}%
  \caption{}
\end{figure}

Starting with the $\text{RCHI}_1$ model which contains a $\phi^{-1}$ correction term, it can be seen in figure \ref{RCHI_1_rns}, that the $1\sigma$ region contains those curves with $50 \lesssim N_k \lesssim 62$. Taking into account reheating consistency in figure \ref{RCHI_1_Tns}, we see that the reheating-consistent curves are those with $58 \lesssim N_k \lesssim 62$. However, these curves only slightly fit into the acceptable region of the $T_{re}-n_s$ plot over the parameter range $1\leq A_I \leq 40$, so the likelihood that they would survive any further constraining by more precise measurements of $n_s$ is low.

\begin{figure}[h]
  \centering
  \begin{subfigure}{.45\linewidth}
    \centering
    \includegraphics[width = \linewidth]{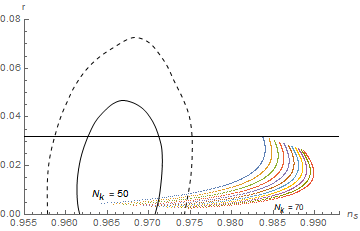}
    \caption{Numerical solutions of the tensor-scalar ratio $r$ and spectral index $n_s$ for the $\text{RCHI}_{10}$ model over the $50 \leq N_k \leq 70$.}
    \label{RCHI_10_rns}
  \end{subfigure}%
  \hspace{3em}%
  \begin{subfigure}{.45\linewidth}
    \centering
    \includegraphics[width = \linewidth]{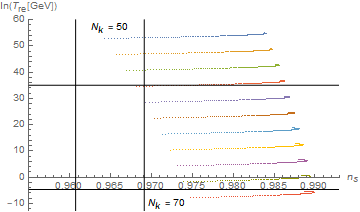}
    \caption{The (log) reheating temperature of the $\text{RCHI}_{10}$ model against the spectral index $n_s$ over $50\leq N_k \leq 70$.}
    \label{RCHI_10_Tns}
  \end{subfigure}%
  \caption{}
\end{figure}

As we increase the power of the correction term through to $p=10$, the effect of this term in the potential is of course much stronger at small $\phi$ where the model exits inflation, and much weaker at large $\phi$ where it starts. Figures for $2\leq p \leq 9$ are given in Appendix C. The effect on the $r-n_s$ plot compared to that of $\text{RCHI}_1$ is to push out of a few of the curves, leaving those with $50 \lesssim N_k \lesssim 60$ in the $1\sigma$ region. This is a relatively small change in the $r-n_s$ parameter space. Conversely, in figure \ref{RCHI_10_Tns} we see that all of the curves have been pushed out of the acceptable in the reheating plot, leaving none that are reheating consistent over this parameter range. 

\begin{figure}[h]
  \centering
  \begin{subfigure}{.45\linewidth}
    \centering
    \includegraphics[width = \linewidth]{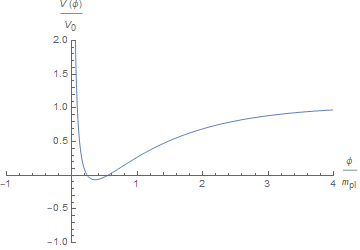}
    \caption{The $\text{RCHI}_1$ potential at $A_I = 1$}
    \label{V_RCHI1}
  \end{subfigure}%
  \hspace{3em}%
  \begin{subfigure}{.45\linewidth}
    \centering
    \includegraphics[width = \linewidth]{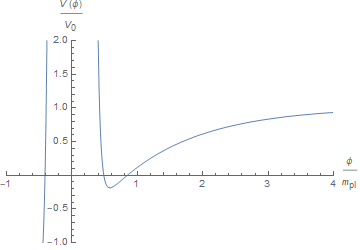}
    \caption{The $\text{RCHI}_{10}$ potential at $A_I = 1$.}
    \label{V_RCHI10}
  \end{subfigure}%
  \caption{}
\end{figure}

Clearly the models become less viable as one increases the strength of the correction term. We also see from figures \ref{V_RCHI1} and \ref{V_RCHI10} that as the the power of the correction term increases the approximation for the coefficient $\alpha_p$ given in equation \ref{a_RCHI} becomes less accurate and the local minima is displaced from $V(\phi_0)\simeq0$. Given that the $\text{RCHI}_1$ model only just survives reheating-consistency, it's likely that RCHI and corrections thereof will not fair well under any future, more precise measurements.

\newpage
\clearpage

\section{Exponential SUSY Inflation}
The final  model we will investigate is that of Exponential SUSY Inflation (ESI). ESI models are governed by potentials of the form 
\begin{equation} 
\label{V_ESI}
    V(\phi) - V_0\left(1-e^{-q\frac{\phi}{m_{pl}}}\right)
\end{equation}
where $q$ is the free parameter of order 1.

Potentials of this form appear in a broad range of literature \cite{Obukhov:1993fd,Stewart:1994ts,Dvali:1998pa,Cicoli:2008gp,Giudice:2010ka}, so we will not discuss the precise nature in which they occur in detail and only focus on the inflationary predictions of the potential \ref{V_ESI} and modifications thereof. Exponential SUSY inflation is a large field inflation model, one can calculate exactly the field value when inflation ends \cite{Martin:2013tda}

\begin{equation}
    \frac{\phi_\textrm{end}}{m_{pl}} = \frac{1}{q}\ln\left(1+\frac{q}{\sqrt{2}}\right)
\end{equation}

For $q\sim\mathcal{O}(1)$, inflation ends at small field values $\phi_{end}/m_{pl}\lesssim 1$, after which the potential becomes negative. The ESI model is another of the single field inflation models featured in \cite{Martin:2013tda} which fits well within the Planck 2018 bounds. However like RCHI and QH, the potential remains unregularized and thus the model cannot lead to any physically realised cosmology.

\begin{figure}[!htb]
  \centering
  \begin{subfigure}{.45\linewidth}
    \centering
    \includegraphics[width = \linewidth]{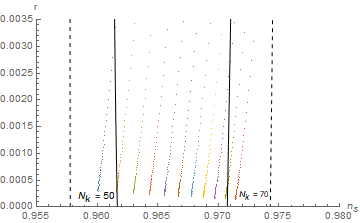}
    \caption{Numerical solutions of the tensor-scalar ratio $r$ and spectral index $n_s$ for the ESI model with $0<q<10$ over $50 \leq N_k \leq 70$. The solid and dashed curves represent the Planck 2018 $1\sigma$ and $2\sigma$ bounds respectively.}
    \label{ESI_rns}
  \end{subfigure}%
  \hspace{3em}%
  \begin{subfigure}{.45\linewidth}
    \centering
    \includegraphics[width = \linewidth]{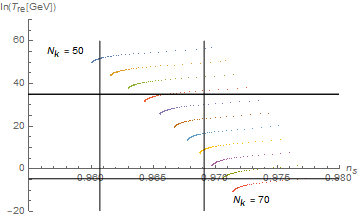}
    \caption{The (log) reheating temperature of the ESI model with against the spectral index $n_s$ with $0<q<10$ over $50\leq N_k \leq 70$. The vertical solid lines represent the Planck 2018 spectral index bounds $0.9607\lesssim n_s \lesssim 0.9691$ and the horizontal solid lines represent the reheating temperature bounds $0.01\text{GeV} \lesssim T_{re} \lesssim 10^{16}\text{GeV}$}
    \label{ESI_Tns}
  \end{subfigure}%
  \caption{}
\end{figure}

If one considers only the $r-n_s$ plot in figure \ref{ESI_rns}, ESI is among the most promising of the three models investigated in this paper, with all curves between $52\lesssim N_k \lesssim 68$ lying within the $1\sigma$ region. In figure \ref{ESI_Tns} however, only the curves for $56 \lesssim N_k \lesssim 64$ lie withing the acceptable region, and much like for the $\text{RCHI}_p$ models, only a small proportion of the curves fit within the Planck 2018 $n_s$ bounds, particularly for larger $N_k$, making it difficult for this model to remain viable if the parameter space were to be further constrained towards the centre of the rectangular bound. Of course we argue that these results are of little significance due to the unregularized behaviour of the potential. Just as with QH and RCHI this can be corrected by taking the square of the potential and calculating the resulting spectral index, tensor fraction and temperature at the end of reheating.

\begin{figure}[h]
  \centering
  \begin{subfigure}{.45\linewidth}
    \centering
    \includegraphics[width = \linewidth]{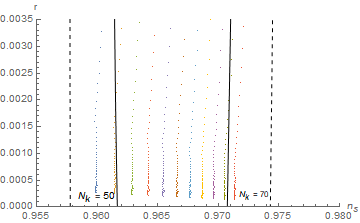}
    \caption{Numerical solutions of the tensor-scalar ratio $r$ and spectral index $n_s$ for the ESI-Squared model with $0<q<10$ over $50 \leq N_k \leq 70$. The solid and dashed curves represent the Planck 2018 $1\sigma$ and $2\sigma$ bounds respectively.}
    \label{ESI_Squared_rns}
  \end{subfigure}%
  \hspace{3em}%
  \begin{subfigure}{.45\linewidth}
    \centering
    \includegraphics[width = \linewidth]{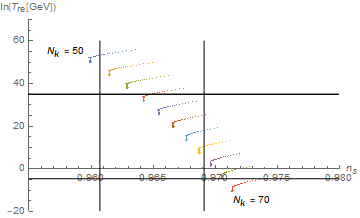}
    \caption{The (log) reheating temperature of the ESI-Squared model with against the spectral index $n_s$ with $0<q<10$ over $50\leq N_k \leq 70$. The vertical solid lines represent the Planck 2018 spectral index bounds $0.9607\lesssim n_s \lesssim 0.9691$ and the horizontal solid lines represent the reheating temperature bounds $0.01\text{GeV} \lesssim T_{re} \lesssim 10^{16}\text{GeV}$}
    \label{ESI_Squared_Tns}
  \end{subfigure}%
  \caption{}
\end{figure}

\newpage

Just as with quartic hilltop and RCHI, one may try simply squaring the ESI potential to obtain a new potential that is bounded below. The squaring of the potential is motivated simply as a means of obtaining a new potential which behaves similarly to the original potential during inflation but has modified behaviour as the inflaton exits inflation. Namely that the potential is bounded from below and has a stable vacuum. From both figures \ref{ESI_Squared_rns} \& \ref{ESI_Squared_Tns} we see that squaring the potential does little to change the inflationary predictions of the model but gives allows the potential to produce a cosmology compatible with a universe that does not immediately collapse after inflation. The reheating-consistent curves remain those with $56\lesssim N_k \lesssim 64$, although they still cover only a small fraction of the parameter space in figure \ref{ESI_Squared_Tns}.

\begin{figure}[h]
  \centering
  \begin{subfigure}{.45\linewidth}
    \centering
    \includegraphics[width = \linewidth]{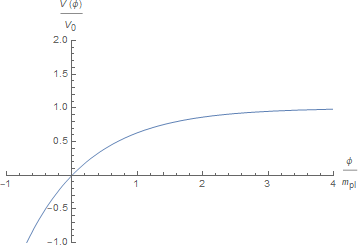}
    \caption{The ESI potential at $q = 1$}
    \label{V_ESIf}
  \end{subfigure}%
  \hspace{3em}%
  \begin{subfigure}{.45\linewidth}
    \centering
    \includegraphics[width = \linewidth]{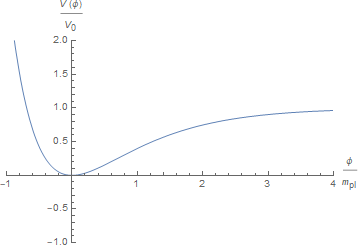}
    \caption{The ESI-Squared potential at $q = 1$.}
    \label{V_ESI_S}
  \end{subfigure}%
  \caption{}
\end{figure}

Unlike QH, we cannot add polynomial terms as an option to correct the ESI model at the end of inflation, since the exponential function will dominate over any $\phi^p$ term we can add towards the end of inflation. We could however, consider adding an inverse power correction of the form $\phi^{-p}$ as we did with RCHI to form the family of $\text{RCHI}_p$ models. 

Consider a potential of the form 
\begin{equation}
V(\phi) = 1 - e^{-q\frac{\phi}{m_{pl}}} + \alpha_p \phi^{-p}
\end{equation}

Just as with the $\text{QH}_p$ and $\text{RCHI}_p$ models, we must adjust the coefficient $\alpha_p$ to ensure that the potentials at the VEV remains at $V(\phi_0) \simeq 0$ to first order so that inflation ends in a finite time. The VEV $\phi_0$ satisfies $V'(\phi) = 0$ and thus
\begin{equation}
\label{ESI_VEV}
    \frac{q}{m_{pl}}e^{-q\frac{\phi_0}{m_{pl}}} - p\alpha_p\phi_0^{-(p+1)} = 0
\end{equation}

For sufficiently small $\phi_0$ equation \ref{ESI_VEV} is dominated by the $\phi_0^{-(p+1)}$ term and thus to first order
\begin{equation}
    \phi_0^{p+1} \simeq \frac{p\alpha_p m_{pl}}{q}
\end{equation}
Demanding that the potential at the VEV is zero, one arrives at the equation
\begin{equation}
    \phi_0\left(1+\frac{1}{p}\right) = 0
\end{equation}
for generic $p$ this would require $\alpha_p = 0$,thus to a first approximation in the methods used here, there are no such correction terms that we can find for ESI. There may be other regularization methods outside the scope of this analysis which could perform this role.

\section{Discussion}
In this paper we have investigated three promising candidate single field inflation models. These models all belong to the subset of candidate inflation models that are capable of producing acceptable an acceptable spectral index and tensor fraction in light of the Planck 2018 data, but do not predict any kind of sensible cosmology that is consistent with what we are simply around to observe the universe today. The issue lies in the unstabilized nature of the potentials after inflation ends. The quartic hilltop, radiatively corrected higgs and exponential SUSY potentials do not posses local minima and so there is no stable vacuum about which the inflaton may oscillate and reheat the universe after expansion. In fact, since the potentials are unbounded from below, if the energy density of the universe continues to be dominated by the inflaton it will collapse on a time scale $t = H^{-1}$ after the end of inflation. This is a generic feature of any such potential that is not stabilized after inflation. The question remains as to whether regularizing the behaviour of these potentials has a significant effect on the inflationary predictions of the models, which indeed it does.

The corrections that are made to the inflationary potentials in this model are not motivated by a particular physical principle, however such potentials do indeed have to be corrected, otherwise the resulting cosmology is simply not compatible with the cosmology we observe today. We show that any such correction terms are significant in terms of their effect on the inflationary predictions of the model. There are of course examples of more physically motivated corrections to inflationary models that previously suffered from the same issue of unboundedness. Brane inflation models featured in \cite{Martin:2013tda} can be consistently modified in a way that does not introduce new fine tuning parameters, to fit the Planck 2018 data in KKTLI inflation \cite{Kallosh:2018zsi}. Likewise, the Higgs inflation model featured also has a physically motivated modification scheme that is consistent with Planck \cite{Ferrara:2010in,Bezrukov:2010jz}. In this approach, the corrections to the functions $F(h)$ and $W(h)$ in equations \ref{F} \& \ref{W} are calculated in the Einstein frame, giving an inflationary potential \ref{RCHI_Reg} that is everywhere positive with a local minima.

Potentials with regularized behaviour may undergo the reheating period after inflation without the universe collapsing in the process. The reheating temperature must at least as large as the BBN energy scale, and no greater than the energy scale of inflation, and so is loosely bounded by $0.01 \text{GeV} \lesssim T_{re} \lesssim 10^{16}\text{GeV}$. Since $T_{re}$ depends explicitly on the amount of inflation that has occurred, through $N_k$ and $H_k$. Demanding reheating-consistency further constrains the acceptable region of the $r-n_s$ parameter space when combined with the Planck 2018 and BICEP/Keck Array measurements. Furthermore, at least approximately 60 efolds of inflation are required to solve the horizon problem \cite{Martin:2003bt}, this consideration also allows us to pinpoint the regions of $r-n_s$ and $T_{re}-n_s$ inhabited by these regularized models.

For the quartic hilltop model investigated in \cite{Hoffmann:2021vty}, before taking into account reheating consistency, one may fit all curves with $53 \lesssim N_k$ into the acceptable region of $r-n_s$ parameter space, however after regularizing to form the QHS model, reheating consistency reduces this to only the curves with $63 \lesssim N_k \lesssim 68$ corresponding to reheating temperatures $1.8 \times 10^{-2} \text{GeV} \lesssim T_{re} \lesssim 6.0 \times 10^4 \text{GeV}$. In this paper we investigate further options for correcting the behaviour of QH by adding polynomial terms $\phi^p$ forming the class of $QH_p$ models for $5\leq p \leq 10$. The lowest order corrected model $QH_5$ has reheating consistent curves $60 \lesssim N_k \lesssim 68$ corresponding to reheating temperatures $1 \text{GeV} \lesssim T_{re} \lesssim 7.2 \times 10^{10}\text{GeV}$. As we increase the strength of the correction term through to $QH_{10}$, the reheating-consistent curves in the $r-n_s$ parameter space still remain only those with $60 \lesssim N_k \lesssim 68$ as the spectral index and tensor fraction $N_k$ efolds before the end of inflation are not particularly sensitive to the strength of the correction term which kicks in towards the end of inflation. However the reheating temperature is sensitive to this change and for $QH_{10}$ these curves correspond to reheating temperatures of $8.2 \times 10^{-2} \text{GeV}\lesssim T_{re} \lesssim 1.1 \times 10^{13}\text{GeV}$

The next model in our discussion is that of Radiatively Corrected Higgs Inflation. Out of the three models considered in this paper, RCHI appears to be the least favourable, having already started in a precarious position even before regularization. Un-regularized RCHI potential allows only those curves with $50\lesssim N_k \lesssim 60$ in the acceptable region of $r-n_s$ parameter space, considering that we require at least 60 efolds for the horizon problem that leaves only a singular curve $N_k \simeq 60$ as a potential candidate. However, none of the curves enter the region of reheating-consistency in the $T_{re}-n_s$ space. As we did with quartic hilltop, we may attempt to regularize this model by simply squaring the potential. The RCHI-Squared potential however only admits curves with $50\lesssim N_k \lesssim 58$ in the acceptable region of $r-n_s$ space, so it does not address the horizon problem, which is an important corner stone of scalar field inflation models. Squaring the RCHI potential significantly alters the shape of the potential during inflation, so we may look for options that only affect the potential towards the end of inflation when the inflation field $\phi$ is small. The simplest such choices would be inverse power corrections of the form $\phi^{-p}$, forming a family of $\text{RCHI}_p$ models. We investigate these for $1\leq p \leq 10$. $\text{RCHI}_1$ admits curves with $58 \lesssim N_k \lesssim 62$ in the acceptable region of $r-n_s$ space, further constrained to $60\lesssim N_k \lesssim 62$ to be consistent with the horizon problem, corresponding to reheating temperatures $2.4\times 10^{7}\text{GeV} \lesssim T_{re} \lesssim 3.6\times 10^{9}\text{GeV}$. This simple correction makes the model very tightly bound and ideal for further analysis under more precise measurements of the spectral index and tensor fraction. As we increase the power of the correction term, there are less reheating-consistent curves available in the $r-n_s$ parameter space. The highest order corrected model we look at, $\text{RCHI}_{10}$ contains no reheating-consistent curves at all.

The final model discussed in this paper is that of exponential SUSY inflation. This potential in its un-regulairzed form already inhabits  promising regions of the $r-n_s$ and $T_{re}-n_s$ parameter spaces. When squared to form a regularized ESI-squared model, the reheating consistent curves are those with $56 \lesssim N_k \lesssim 64$. Taking into account the horizon problem this is further reduced to $60 \lesssim N_k \lesssim 64$, a corresponding bound on the reheating temperature of $1.5 \times 10^2 \text{GeV}\lesssim T_{re} \lesssim 4.9 \times 10^8 \text{GeV}$. Unlike QH, adding polynomial corrections to ESI does not stabilise the potential as the exponential function dominates any such term that we could add. We show also that is it not possible to add inverse power terms that can be investigated analytically to first order.

Single field inflation models provide a versatile landscape of models to approach the problem of explaining cosmological inflation, but one must take great care that any model under consideration produces a cosmology after inflation that is consistent with what we observe today. In particular, it is important that the inflationary models we consider in this paper have stable vacuums which prevent the universe from collapsing shortly after inflation ends. This is an important consideration for as we have shown in the cases of quartic hilltop and radiatively corrected Higgs inflation, how one decides to regularize the potential can have significant effects on the observable that we measure today and may determine the viability of a given model when compared to observational data. In this work we make use of polynomial and inverse power correction terms, it is certainly worth noting that for a scalar field theory in $d=4$ dimensions, only $\phi^4$ interaction terms are renormalizable \cite{Li:2012mwa,Ball:1993zy}. So even models such as these with corrected potentials cannot be quantized to form self-consistent QFT's, they can only be considered as classical effective field theories for computing inflationary observables.

\clearpage
\newpage

\appendix
\section{Numerical Simulations}
Throughout this work we make frequent use of numerical simulations to calculate the spectral index, tensor fraction and temperate at the end of reheating. These calculations are performed using a bespoke Mathematica notebook available at \url{https://github.com/JoshHoffmann/SFSRIA}. This code allows for the analysis of generic single field slow-roll inflation models by numerical integrating the exact equations of motion \ref{EoM} at a fixed number of remaining efolds of inflation $N_k$, over parameter ranges and suitable initial conditions set by the user. Integrating the equations of motion up to the end of inflation allows one to extract the spectral index $n_s$ and tensor fraction $r$ as well as the energy density of the scalar field at the end of inflation and Hubble parameter when the picot scale exits the horizon $H_k$. Thus we are also able to calculate the reheating temperature and number of reheating efolds given by equations \ref{T_re} \& \ref{N_re} respectively.

\section{$\text{QH}_p$ Correction figures}
\begin{figure}[h]
  \centering
  \begin{subfigure}{.45\linewidth}
    \centering
    \includegraphics[width = \linewidth]{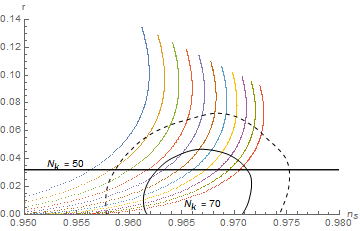}
    \caption{Numerical solutions of the tensor-scalar ratio $r$ and spectral index $n_s$ for the $\text{QH}_6$ model over the $50 \leq N_k \leq 70$.}
  \end{subfigure}%
  \hspace{3em}%
  \begin{subfigure}{.45\linewidth}
    \centering
    \includegraphics[width = \linewidth]{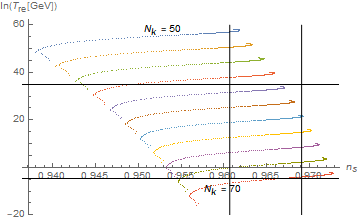}
    \caption{The (log) reheating temperature of the $\text{QH}_6$ model against the spectral index $n_s$ over $50\leq N_k \leq 70$.}
  \end{subfigure}%
  \caption{}
\end{figure}

\begin{figure}[h]
  \centering
  \begin{subfigure}{.45\linewidth}
    \centering
    \includegraphics[width = \linewidth]{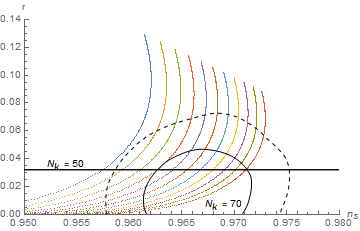}
    \caption{Numerical solutions of the tensor-scalar ratio $r$ and spectral index $n_s$ for the $\text{QH}_7$ model over the $50 \leq N_k \leq 70$.}
  \end{subfigure}%
  \hspace{3em}%
  \begin{subfigure}{.45\linewidth}
    \centering
    \includegraphics[width = \linewidth]{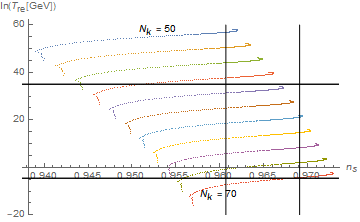}
    \caption{The (log) reheating temperature of the $\text{QH}_7$ model against the spectral index $n_s$ over $50\leq N_k \leq 70$.}
  \end{subfigure}%
  \caption{}
\end{figure}
\newpage
\clearpage

\begin{figure}[h]
  \centering
  \begin{subfigure}{.45\linewidth}
    \centering
    \includegraphics[width = \linewidth]{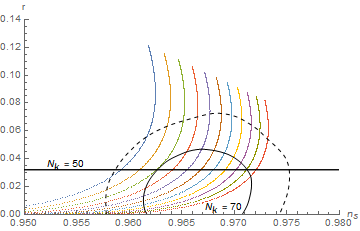}
    \caption{Numerical solutions of the tensor-scalar ratio $r$ and spectral index $n_s$ for the $\text{QH}_9$ model over the $50 \leq N_k \leq 70$.}
  \end{subfigure}%
  \hspace{3em}%
  \begin{subfigure}{.45\linewidth}
    \centering
    \includegraphics[width = \linewidth]{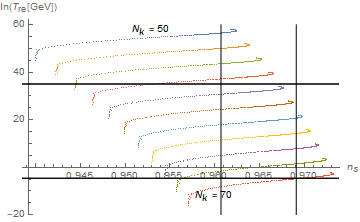}
    \caption{The (log) reheating temperature of the $\text{QH}_9$ model against the spectral index $n_s$ over $50\leq N_k \leq 70$.}
  \end{subfigure}%
  \caption{}
\end{figure}

\section{$\text{RCHI}_p$ Correction Figures}
\begin{figure}[!htb]
  \centering
  \begin{subfigure}{.45\linewidth}
    \centering
    \includegraphics[width = \linewidth]{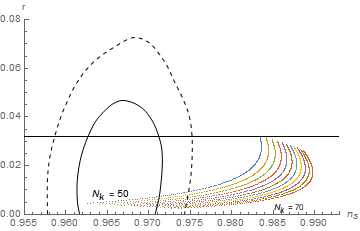}
    \caption{Numerical solutions of the tensor-scalar ratio $r$ and spectral index $n_s$ for the $\text{RCHI}_2$ model over the $50 \leq N_k \leq 70$.}
  \end{subfigure}%
  \hspace{3em}%
  \begin{subfigure}{.45\linewidth}
    \centering
    \includegraphics[width = \linewidth]{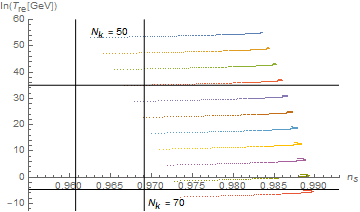}
    \caption{The (log) reheating temperature of the $\text{RCHI}_2$ model against the spectral index $n_s$ over $50\leq N_k \leq 70$.}
  \end{subfigure}%
  \caption{}
\end{figure}

\begin{figure}[h]
  \centering
  \begin{subfigure}{.45\linewidth}
    \centering
    \includegraphics[width = \linewidth]{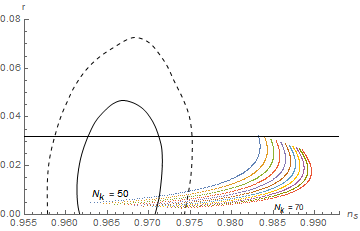}
    \caption{Numerical solutions of the tensor-scalar ratio $r$ and spectral index $n_s$ for the $\text{RCHI}_3$ model over the $50 \leq N_k \leq 70$.}
  \end{subfigure}%
  \hspace{3em}%
  \begin{subfigure}{.45\linewidth}
    \centering
    \includegraphics[width = \linewidth]{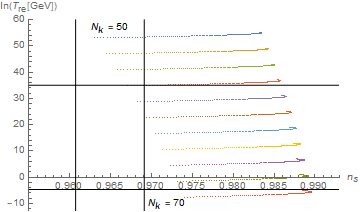}
    \caption{The (log) reheating temperature of the $\text{RCHI}_3$ model against the spectral index $n_s$ over $50\leq N_k \leq 70$.}
  \end{subfigure}%
  \caption{}
\end{figure}

\begin{figure}[h]
  \centering
  \begin{subfigure}{.45\linewidth}
    \centering
    \includegraphics[width = \linewidth]{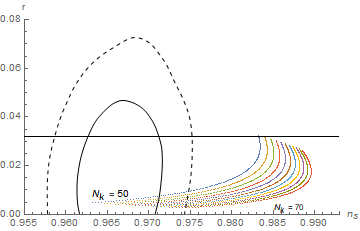}
    \caption{Numerical solutions of the tensor-scalar ratio $r$ and spectral index $n_s$ for the $\text{RCHI}_4$ model over the $50 \leq N_k \leq 70$.}
  \end{subfigure}%
  \hspace{3em}%
  \begin{subfigure}{.45\linewidth}
    \centering
    \includegraphics[width = \linewidth]{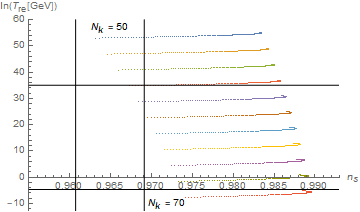}
    \caption{The (log) reheating temperature of the $\text{RCHI}_4$ model against the spectral index $n_s$ over $50\leq N_k \leq 70$.}
  \end{subfigure}%
  \caption{}
\end{figure}

\begin{figure}[h]
  \centering
  \begin{subfigure}{.45\linewidth}
    \centering
    \includegraphics[width = \linewidth]{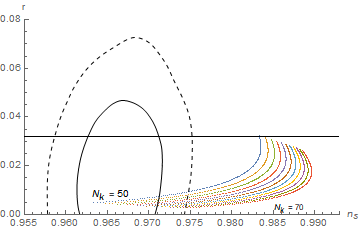}
    \caption{Numerical solutions of the tensor-scalar ratio $r$ and spectral index $n_s$ for the $\text{RCHI}_5$ model over the $50 \leq N_k \leq 70$.}
  \end{subfigure}%
  \hspace{3em}%
  \begin{subfigure}{.45\linewidth}
    \centering
    \includegraphics[width = \linewidth]{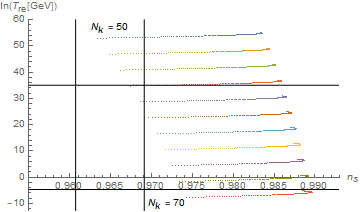}
    \caption{The (log) reheating temperature of the $\text{RCHI}_5$ model against the spectral index $n_s$ over $50\leq N_k \leq 70$.}
  \end{subfigure}%
  \caption{}
\end{figure}

\begin{figure}[h]
  \centering
  \begin{subfigure}{.45\linewidth}
    \centering
    \includegraphics[width = \linewidth]{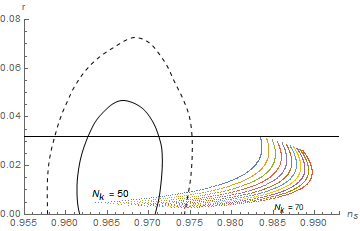}
    \caption{Numerical solutions of the tensor-scalar ratio $r$ and spectral index $n_s$ for the $\text{RCHI}_6$ model over the $50 \leq N_k \leq 70$.}
  \end{subfigure}%
  \hspace{3em}%
  \begin{subfigure}{.45\linewidth}
    \centering
    \includegraphics[width = \linewidth]{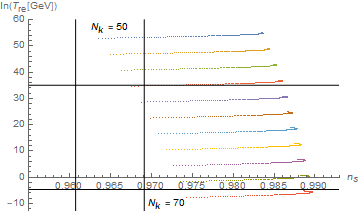}
    \caption{The (log) reheating temperature of the $\text{RCHI}_6$ model against the spectral index $n_s$ over $50\leq N_k \leq 70$.}
  \end{subfigure}%
  \caption{}
\end{figure}

\begin{figure}[h]
  \centering
  \begin{subfigure}{.45\linewidth}
    \centering
    \includegraphics[width = \linewidth]{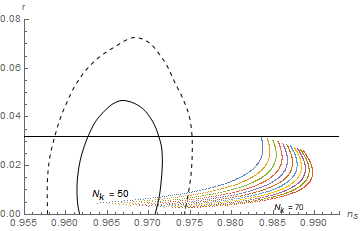}
    \caption{Numerical solutions of the tensor-scalar ratio $r$ and spectral index $n_s$ for the $\text{RCHI}_7$ model over the $50 \leq N_k \leq 70$.}
  \end{subfigure}%
  \hspace{3em}%
  \begin{subfigure}{.45\linewidth}
    \centering
    \includegraphics[width = \linewidth]{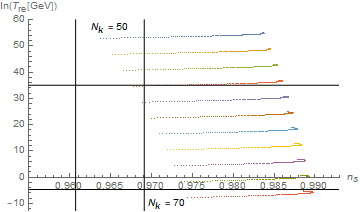}
    \caption{The (log) reheating temperature of the $\text{RCHI}_7$ model against the spectral index $n_s$ over $50\leq N_k \leq 70$.}
  \end{subfigure}%
  \caption{}
\end{figure}

\begin{figure}[h]
  \centering
  \begin{subfigure}{.45\linewidth}
    \centering
    \includegraphics[width = \linewidth]{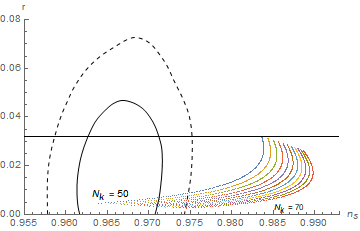}
    \caption{Numerical solutions of the tensor-scalar ratio $r$ and spectral index $n_s$ for the $\text{RCHI}_8$ model over the $50 \leq N_k \leq 70$.}
  \end{subfigure}%
  \hspace{3em}%
  \begin{subfigure}{.45\linewidth}
    \centering
    \includegraphics[width = \linewidth]{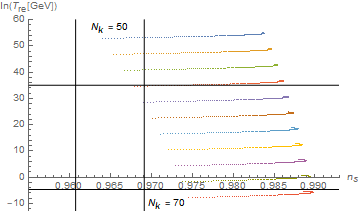}
    \caption{The (log) reheating temperature of the $\text{RCHI}_8$ model against the spectral index $n_s$ over $50\leq N_k \leq 70$.}
  \end{subfigure}%
  \caption{}
\end{figure}
\newpage
\clearpage

\begin{figure}[h]
  \centering
  \begin{subfigure}{.45\linewidth}
    \centering
    \includegraphics[width = \linewidth]{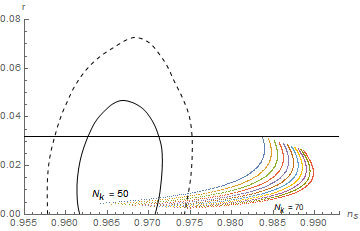}
    \caption{Numerical solutions of the tensor-scalar ratio $r$ and spectral index $n_s$ for the $\text{RCHI}_9$ model over the $50 \leq N_k \leq 70$.}
  \end{subfigure}%
  \hspace{3em}%
  \begin{subfigure}{.45\linewidth}
    \centering
    \includegraphics[width = \linewidth]{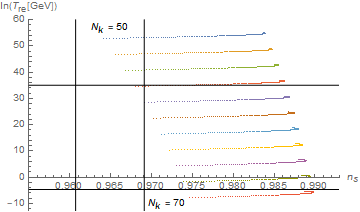}
    \caption{The (log) reheating temperature of the $\text{RCHI}_9$ model against the spectral index $n_s$ over $50\leq N_k \leq 70$.}
  \end{subfigure}%
  \caption{}
\end{figure}

\bibliography{refs.bib}

\bibliographystyle{ieeetr}

\end{document}